\documentstyle[prb,aps,floats,epsfig]{revtex}

\begin{document}
\draft
\title{Possible explanation for the absence of bilayer splitting in YBCO}

\author{
  H. Monien, N. Elstner and A. J. Millis \cite{Baltimore}
}
\address{
  Physikalisches Institut, 
  Universit\"at Bonn, \\
  Nu\ss allee 12,\\ 
  D-53115 Bonn,\\ 
  Germany 
}
\twocolumn[
\date{\today}
\maketitle
\widetext
\begin{abstract}
  \begin {center}
    \parbox{14cm}{ 
      It has been claimed that the absence of the bilayer splitting
      in the high-temperature superconductor YBCO is a strong experimental
      indication that there are no coherent quasiparticles present in the CuO
      planes \cite{PWAnderson91}. We study a pair of strongly correlated 
      planes which are connected by a hopping transfer integral $t_{\perp}$
      in the limit of large in-plane coordination number. The effect of 
      the correlation is incorporated in a dynamical mean-field theory, 
      where the Weiss field is determined by a two-site Hubbard Hamiltonian. 
      We have solved this problem by numerical techniques and present results 
      for the spectral function $\rho$ for relevant parameters of the model. 
      For small $t_{\perp}$ we find the coherent hoping between planes 
      to be proportional to the in-plane quasiparticle renormalisation. }
  \end{center}
\end{abstract}

\pacs{\hspace{1.9cm}PACS: 71.10.-w,71.10.Ay,71.10.Fd,71.28.+d,71.30.+h,74.25.Jb,74.72.-h}
]
\narrowtext
\vfill\eject

The transport properties of the high temperature superconductors show an
unusual behavior in many respects. Anderson \cite{PWAnderson91} has emphasized
the difference between the transport properties in a CuO plane and
perpendicular to the CuO plane stressing in particular the optical
conductivity, in the c-direction perpendicular to the CuO planes
\cite{Timusk93}.  Also photoemission experiments \cite{Shen95,Ding96} in YBCO
show only a small if any splitting of the bonding and antibonding CuO plane
bands due to coherent hopping between the CuO planes.  This is even more
surprising in view of the relatively large value of the bare hopping matrix
element, $t_\perp \approx 0.3 \, \text{eV}$ compared to $t_\parallel \approx 1
\, \text{eV}$, estimated from bandstructure calculations \cite{Pickett89}. The
experimental data have been interpreted as a signal for the absence of coherent
quasiholes in the CuO plane \cite{PWAnderson91}.

The question of how interactions affect the coherent transport between the
planes has been studied mostly on one dimensional strongly correlated systems
because in that case it is clear that the excitations above the ground state
are spinons and holons as in the original proposal by Anderson. Two coupled
chains (ladders) have been recently studied in detail by many groups with
numerical and analytic techniques
\cite{Strong92,Clarke94,Rice93,Noack94,TwoChainsReview,Tsunetsugu97}. The
result was that the hopping matrix element $t_\perp$ is a relevant
perturbation for physically relevant values of the Luttinger liquid exponents
and that the transport between the chains is coherent.  Recently it has
been claimed from an exact diagonalization study that for longer range
interactions in the chains the coherent hopping between the chains can be
reduced to zero for a finite value of the bare hopping matrix element
$t_\perp$ \cite{Poilblanc96}.

The problem of two coupled CuO planes is clearly more difficult since
at this point the ground state and the excitation spectrum even for a
single plane of the two dimensional Hubbard model away from half
filling are not known. The problem has been studied with numerical
techniques like Quantum Monte Carlo (QMC) \cite{Scalettar94} and the
fluctuation exchange approximation (FLEX) approximation
\cite{Putz96,Grabowski97,Dahm97}. The subtle features of the dynamics of the
quasiparticles in the coupled planes can currently not be resolved in
the QMC due to finite size effects although the results for one plane
look encouraging \cite{Noack94}. The FLEX calculations
\cite{Putz96,Grabowski97,Dahm97} are able to resolve the dynamics and yield
surprisingly good results when compared to the QMC calculation but this 
approximation is uncontrolled in the regime where the interaction is large.

\begin{figure}[h]
\protect\centerline{\epsfig{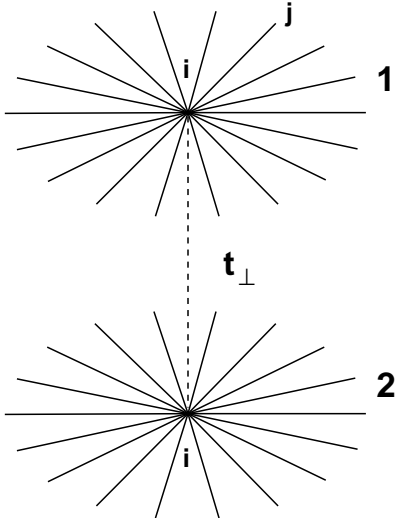}}
\protect\caption{
Site of the two planes connected by nonzero hopping matrix elements
}
\label{fig:1}
\end{figure}

In this paper we present a calculation of the spectral function and the
reduction of the coherent hopping in an approximation where the local
correlations are taken into account exactly.  We are in particular interested
in the dynamics of the quasiparticles. We consider a model of two strongly
correlated planes. We neglect the Coulomb interaction between the two planes.
The motion of the holes in each plane is described by a Hubbard
Hamiltonian.
\begin{equation}
  H_a = - t \sum\limits_{<ij>\sigma} 
  c^{\dagger}_{ai \sigma}c^{\phantom{\dagger}}_{aj \sigma} 
  - \mu \sum_{i\sigma} n_{ai\sigma} +
  U \sum\limits_i n_{ai \uparrow} n_{ai \downarrow},
\end{equation}
where $t$ is the hopping matrix element between two nearest neighbors $i$ and
$j$, $\mu$ is the chemical potential, $c^\dagger_{ai\sigma}$ is the electron
creation operator on site $i$ with spin $\sigma$ in plane $a$ ($a=1,2$) and
$n_{ai\sigma}$ is the electron number operator for spin $\sigma$ at site $i$
in plane $a$, $n_{ai \sigma} = c^\dagger_{ai\sigma}
c^{\phantom{\dagger}}_{ai\sigma}$.  We denote the electron creation and
annihilation operators in plane 1 by $c^\dagger_{1i}$ and
$c^{\phantom{\dagger}}_{1i}$ respectively and in plane 2 by $c^\dagger_{2i}$
and $c^{\phantom{\dagger}}_{2i}$.  The two planes are coupled by the hopping
term
\begin{equation}
  H_{12} = - t_\perp \sum_{i \sigma} 
  \left( c^{\dagger}_{1i \sigma} c^{\phantom{\dagger}}_{2i \sigma}
    + c^{\dagger}_{2i \sigma} c^{\phantom{\dagger}}_{1i \sigma} 
  \right)
\end{equation}
with the hopping matrix element $t_\perp$ connecting site $i$ in plane
$1$ with site $i$ in plane $2$. The complete Hamiltonian for the
bilayer is then $H = H_1 +H_2 + H_{12}$. An exact solution of this
problem at this point in time is clearly impossible since already
$H_1$ and $H_2$ cannot be treated exactly in two spatial dimensions.
However some progress can be made by introducing the so called dynamical 
mean field approximation in which the local correlations are treated exactly.
For a review of this method see e.g. \cite{Freericks94,Georges96}.
The important point is that the dynamics of the quasiparticles hopping
onto a given site and leaving a given site is kept.
To apply this approximation to the problem at hand the number of
neighbors $z$ of each site in a given plane $(1, 2)$ to its neighbors
in the {\em same} plane has to go to infinity, see Fig.~\ref{fig:1}.
In this approach the phase information of the electrons hopping from a
site $j$ to a given site $i$ in the same plane is lost because the
phase information is averaged of over the large number of sites $j$
connected to $i$. The hopping term between the planes has to be treated
separately because there is only one site in plane $2$ connected to a
given site in the plane $1$.

The non interacting $(U=0)$ Green's function $\bbox{G}^{-1}_0$ is defined as
$\left[\bbox{G}_0(k,\tau)\right]_{aa'} = -<T_\tau
c^{\phantom{\dagger}}_{ak\alpha}(\tau) c^{\dagger}_{a'k\beta}(0)>$ with $a, a'
\in$ $\{1, 2\}$.  It is given by
\begin{equation}
  \bbox{G}^{-1}_0(k, \omega ) = \omega - \bbox{H} (U= 0) = 
  \left[ \begin{array}{cc}
      \omega - \epsilon_k & t_\perp \\
      t_\perp & \omega - \epsilon_k
    \end{array} \right]
\end{equation}
In the paramagnetic phase the self energy $\bbox{\Sigma}(k, \omega)$ 
has only two independent components. 
We denote them by $\Sigma_{\parallel}$ and $\Sigma_{\perp}$ 
in the present basis 
\begin{equation}
  \label{eq:Sigma}
  \bbox{\Sigma} = \left[
    \begin{array}{cc}
      \Sigma_\parallel & \Sigma_\perp \\
      \Sigma_\perp & \Sigma_\parallel
    \end{array}
  \right]
\end{equation}
The Hamiltonian is diagonal in the bonding - antibonding representation. The
full Green's function can be obtained from the Dyson equation
\begin{equation}
  \label{eq:Dyson}
  \bbox{G}^{-1}(k, \omega ) = 
  \bbox{G}^{-1}_0(k, \omega ) - \bbox{\Sigma}(k, \omega )
\end{equation}
We are now in the position to derive the dynamical mean field
equations for the coupled planes.  In the limit of large coordination
number the self energy becomes local and therefore $k$-independent
\cite{MuellerHartmann89}.  The only $k$-dependence of the Green's
function is left in the free electron dispersion $\epsilon_k$.  This
$k$-dependence can be absorbed in the density of states 
$D(\epsilon) = \sum_k \delta (\epsilon - \epsilon_k)$,
which of course depends on the type of lattice in which the electrons 
are moving. For the hypercubic lattice $D(\epsilon) = exp({- (\epsilon/\sqrt{2}t)^2})/t\sqrt{2\pi}$. 
The full {\em local} Green's function is therefore given by:
\begin{equation}
  \label{eq:G_local}
  \bbox{G} (\omega) = \int d\epsilon_{\bf k} D(\epsilon_{\bf k}) \left( 
    \bbox{\cal G}^{-1}_0( \epsilon, \omega ) - \bbox{\bbox{\Sigma}}( \omega ) 
  \right)^{-1}
\end{equation}

In the dynamical mean field approach $\bbox{\cal G}^{-1}_0(\tau)$ plays 
the role of a Weiss field. The main point of this approach is that the 
{\em local} self energy of the full problem is determined by the local
impurity model \cite{MetznerVollhardt89}: in our case a two
impurity model. It has the action
\begin{eqnarray}
  \label{eq:S_impurity}
  S = \int^\beta_0 d\tau &&\int^\beta_0 d \tau^\prime
  \sum_\sigma\bar{\psi}_{\sigma}( \tau ) 
  \bbox{\cal G}^{-1}_{0}( \tau-\tau^\prime )
  \psi_{\sigma}( \tau^\prime ) \cr + &&U \sum_{a=1,2} \int^\beta_0 d\tau\;  
  \bar{\psi}_{a\uparrow}( \tau ) \psi_{a\uparrow}( \tau )
  \bar{\psi}_{a\downarrow}( \tau ) \psi_{a\downarrow}( \tau )
\end{eqnarray}
where $\bar{\psi}_\sigma = (\bar{\psi}_{1\sigma}, \bar{\psi}_{2\sigma})$ is a
spinor of Grassmann variables.  We only consider the paramagnetic
state. The Weiss field $\bbox{\cal G}_0^{-1}(\omega)$ is therefore diagonal in
spin space.  Using Eq. \ref{eq:G_local} we obtain the full local Green's
function in the bonding, anti-bonding basis:
\begin{equation}
  \bbox{G}(\omega ) =  
  \left[ \begin{array}{cc}
      {\tilde D}_{+}(\omega) & 0 \\\
      0 & {\tilde D}_{-}(\omega) 
    \end{array} \right]
\end{equation}
where ${\tilde D}_{\pm}(\omega) = 
{\tilde D}(\omega-\Sigma_\parallel(\omega)\pm(t_\perp-\Sigma_\perp(\omega)))$
is the Green's function of the anti-bonding and bonding band respectively and
\begin{equation}
  \label{eq:HilbertTrafo}
  {\tilde D}(\omega) = \int d\epsilon \frac{D(\epsilon)}{\epsilon - \omega}
\end{equation}
is the Hilbert transform of the density of states $D(\epsilon)$.

We solve the two impurity model, Eq. \ref{eq:S_impurity}, using the
''iterated perturbation theory'' (IPT) \cite{Georges92} which
has been shown to yield qualitative agreement with QMC and exact
diagonalization approaches for interaction strength up to $U/t \sim
4$. The self energy is approximated by the first two terms of a
perturbation expansion in the interaction $U$.  However these two
terms are calculated with the local Green's function ${\cal G}_0$ which is
then determined selfconsistently. For doping levels away from half
filling we used an extension of the IPT due to Kajueter and Kotliar
\cite{Kajueter96}.  

\begin{figure}[h]
  \protect\centerline{\epsfig{file=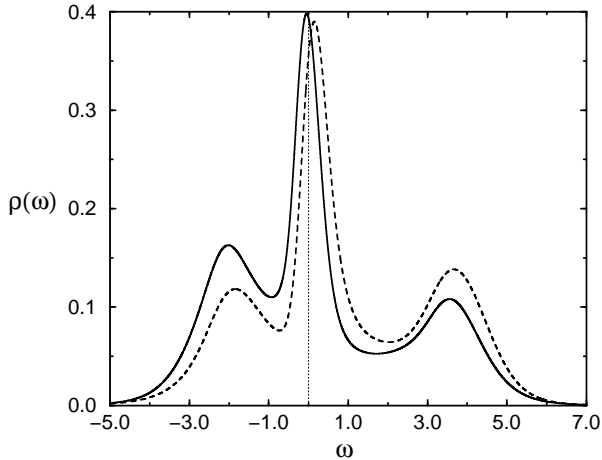, height=7cm}}
  \protect\caption{ Spectral functions for the bonding (solid line)
    and anti-bonding (dashed line) band. The parameters are $U=4$,
    $n=0.4375$ and $t_\perp=0.3$. The dotted vertical line marks the 
Fermi energy.}
\label{fig:spectral-function}
\end{figure}

In Fig.~\ref{fig:spectral-function} the spectral functions of the
bonding and antibonding band are shown as a function of frequency for
a $U=4$, a bilayer hopping of $t_\perp=0.3$ and a filling of $n=0.4375$.
All energies are measured in units of the in-plane hopping matrix
element $t$.  
For each band the lower and upper Hubbard band can be clearly identified.  The
quasiparticle peak close to $\omega=0$ for each band is slightly shifted away
from zero as expected from the noninteracting limit. The first interesting
observation is that the spectral weight in the lower and the upper Hubbard
band is not symmetric anymore even at half filling. The spectral density of
the upper Hubbard band of the bonding band and the lower Hubbard band of the
anti-bonding band are suppressed. With increasing hopping between the planes
the quasiparticle peaks move further apart and the features associated with
the Hubbard bands become less and less pronounced. For very large values of
$t_\perp$ the noninteracting density of states of the two bands is recovered.
The small features in the spectral function which are associated with the
formation of so called shadow bands as obtained in the FLEX calculation
\cite{Grabowski97} can not be found in our calculation, because the finite
momentum antiferromagnetic fluctuations are missing from our calculation.
However the redistribution of weight from the Hubbard bands to the
quasiparticle peak and vice versa has not been resolved in the FLEX perhaps
because it is inherently a weak coupling approach.  At half filling both bands
are filled up to $\omega=0$.  The self energy of the bonding and antibonding
propagator is given by $\Sigma_\parallel(\omega)\pm\Sigma_\perp(\omega)$.  For
small values of $t_\perp$ the quasiparticle weight is strongly reduced.

\begin{figure}[t]
  \protect\centerline{\epsfig{file=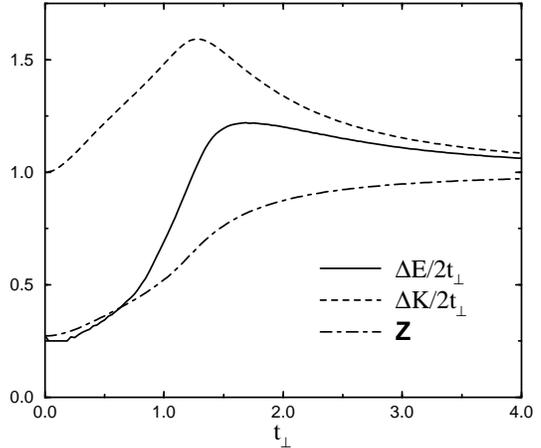,height=7cm}}
  \protect\caption{The splitting $\Delta E / 2 t_\perp$ ($\Delta K / 2
    t_\perp$) of the coherent hopping between planes as a function of
    $t_\perp$. The parameters are $U=4$ and $n=0.5$. The dashed line refers to
    the splitting in energy, the solid line to the splitting in $\epsilon_k$.
    The spectral weight of the quasiparticle peaks is shown as dashed-dotted
    line.}
  \label{fig:reduction}
\end{figure}
Next we discuss the reduction of the coherent hopping between the planes. 
There are two measures of coherent hopping: one is $\Delta E$, the splitting 
in frequency between the quasiparticle peaks in the bonding and antibonding
bands; the other is $\Delta K = 2(t_{\perp} - \Sigma_{\perp}(\omega=0))$, 
the splitting in momentum space between the bonding and antibonding bands at
the Fermi surface. We consider $\Delta E/2t_{\perp}$ first. Its 
$t_{\perp}$ dependence is shown in Fig.~\ref{fig:reduction}. In the 
noninteracting case $\Delta E$ assumes the value $2 t_\perp$.  For small 
$t_\perp$ the ratio $\Delta E/2 t_\perp$ is clearly reduced from the 
noninteracting value of 1. However unless one drives
the system through the Mott transition at very large values of $U$ where the
planes themselves are insulating the reduction factor is nonzero. In the
$t_\perp\rightarrow 0$ limit the reduction factor can be analytically related 
to the reduction of the bandwith in the plane. The off-diagonal part of the
self-energy $\Sigma_\perp$ is proportional to $t_\perp^3$ for small $t_\perp$.
The position of the quasiparticle-peak is given by the poles of ${\tilde
  D}_{\pm}(z)$ in the complex plane.  For small $t_\perp$ and doping not too
far from half filling the position of the pole is given by the zero of $\omega
- {\rm Re}\;\Sigma_\parallel(\omega) \pm t_\perp + O(t_\perp^3) = 0$.  The
selfenergy does not depend on the momentum.  The reduction of the bandwidth is
therefore solely given by
$(1-\partial\Sigma_\parallel(\omega)/\partial\omega)^{-1}$ which is the same
as the reduction of the bandwith due to interaction in a single plane. 
Interestingly the coherent hopping is seen to be enhanced for
larger values of $t_\perp$. This effect can be understood in terms of a
level repulsion of the two site Hubbard model in the regime where $t_\perp \gg
U \gg t$.  For very large values of $t_\perp$ the splitting asymptotically
approaches the value $2t_\perp$ as one would expect from the noninteracting
system.  

\begin{figure}[t]
  \protect\centerline{\epsfig{file=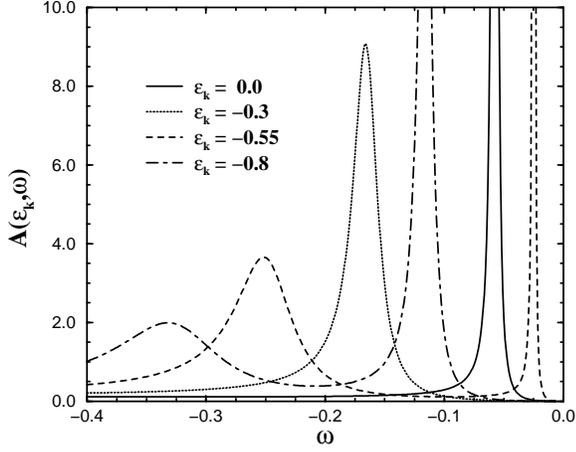,height=7cm}}
  \protect\caption{
    Single particle spectral function $A({\bf \epsilon_k},\omega)$ vs.
    $\omega$ for different values of $\epsilon_{\bf k}$.  The parameters are
    $t_{\perp} = 0.3$, $U=4$ and $n=0.4375$.}
  \label{fig:A_k_omega}
\end{figure}
In Fig.~\ref{fig:A_k_omega} we plotted the spectral density
$A(\epsilon_k,\omega)$ at zero temperature as function of frequency for
parameter values $t_{\perp} = 0.3$ and $U=4$. The wavevector dependence is
only implicitly contained via $\epsilon_k$, because of the local nature of
dynamical mean field theory. For $\epsilon_k$ close to zero we observe one
sharp peak associated with the bonding band just below the Fermi level as one
would expect (The bonding band actually crosses the Fermi level at 
$\epsilon_{\bf k} \approx 0.15 $). 
 When going to lower energies this maximum moves downwards in the
spectrum and very rapidly becomes broader. The broadening is due to strong
correlations which manifest themself in a rapid increase of the self energy 
away from $\omega =0$.  For $\epsilon_k \approx -0.55$ a new sharp peak 
due to the antibonding band appears. The peak of the bonding band, however, 
is already rather broad at this energy. The splitting $\Delta K$ between 
the $\epsilon_{\bf k}$ values where the bonding and antibonding bands cross
the Fermi surface is thus $\Delta K \approx 0.7$ so for these parameters
$\Delta K/ 2t_{\perp} \approx 1.1$.

In conclusion we have studied the problem of two coupled planes in
the dynamical mean field theory in the paramagnetic state as a
function of interaction, doping and interplane hopping matrix
elements. Within our approximation the coherent hopping between the
planes is strongly reduced due to the interaction however it stays
nonzero as long as the planes stay in the paramagnetic phase. It is
still an open question if short range antiferromagnetic correlations
which are completely neglected in our approach can suppress the
coherence between coupled planes even further.  For intermediate
values of the interplane coupling the coherent hopping motion of 
the quasiparticles between planes is increased by the interaction.  
The spectral weight in the Hubbard bands is redistributed in an
unsymmetric way even at half filling. Our approach can be easily
extended to include more realistic density of states. It is
interesting to examine the stability of other groundstates like for
example the antiferromagnetic groundstate.

We would like to acknowledge useful discussions on this problem with 
A.~J.~Ruckenstein, A.~Georges, G.~Kotliar, J.~Schmalian, D.~Poilblanc,
J.~R.~Schrieffer and D. Vollhardt. We are particular indebted to G. Uhrig for
pointing out an error in our early analysis.  We acknowledge the hospitality
of Lucent Technology, the Aspen Center of Physics and the Johns Hopkins
University were part of this work was carried out.  This work was supported in
part by the ISI foundation under grant ERBCHRC-CT920020 and NATO grant
CGR 960680.

\end{document}